# Mechanisms and Geochemical Models of Core Formation

David C. Rubie[1*] and Seth A. Jacobson[1,2]

[1]Bayerisches Geoinstitut, University of Bayreuth, D-95490 Bayreuth, Germany (dave.rubie@uni-bayreuth.de)
[2]Observatoire de la Côte d'Azur, F-06304 Cedex 4 Nice, France
* Corresponding author

**Key points**:
- Metal-silicate segregation during core formation occurred in impact-induced magma oceans
- Single stage core formation models have been superseded by continuous and multistage models
- Coupled dynamical accretion and multi-stage core formation models are now possible

**Abstract**

The formation of the Earth's core is a consequence of planetary accretion and processes in the Earth's interior. The mechanical process of planetary differentiation is likely to occur in large, if not global, magma oceans created by the collisions of planetary embryos. Metal-silicate segregation in magma oceans occurs rapidly and efficiently unlike grain scale percolation according to laboratory experiments and calculations. Geochemical models of the core formation process as planetary accretion proceeds are becoming increasingly realistic. Single stage and continuous core formation models have evolved into multi-stage models that are couple to the output of dynamical models of the giant impact phase of planet formation. The models that are most successful in matching the chemical composition of the Earth's mantle, based on experimentally-derived element partition coefficients, show that the temperature and pressure of metal-silicate equilibration must increase as a function of time and mass accreted and so must the oxygen fugacity of the equilibrating material. The latter can occur if silicon partitions into the core and through the late delivery of oxidized material. Coupled dynamical accretion and multi-stage core formation models predict the evolving mantle and core compositions of all the terrestrial planets simultaneously and also place strong constraints on the bulk compositions and oxidation states of primitive bodies in the protoplanetary disk.

**Index terms: 1015, 1025, 1060, 3610,**

# 1. Introduction

The terrestrial planets, Mercury, Venus, Earth, Moon, and Mars, and at least some much smaller bodies in the asteroid belt (e.g. 4 Vesta), have metallic cores that are surrounded by silicate mantles. Core-mantle structures result from gravity-driven differentiation events that occurred during the early (~100 My) history of the Solar System. During planetary accretion, Fe-rich metal was delivered either in the form of cores of differentiated bodies (as represented by iron meteorites) or as metal that was finely dispersed in a silicate matrix (as represented by chondritic meteorites). In both cases, given the dimensions of planetary mantles, the process of core-mantle differentiation required metal to segregate from silicate over large length scales (e.g. up to 3000 km in the case of the Earth).

Here we review the mechanisms by which metal and silicate segregate to form the cores and mantles of planetary bodies. In addition, we review geochemical models of core formation and consider the implications of these for the evolution of mantle and core chemistries. Some aspects are dealt with briefly in this short review and additional sources of information are provided by Stevenson (1990), Nimmo and Kleine (2015) and Rubie et al. (2003, 2007, 2015a).

# 2. Mechanisms of metal-silicate segregation

For metal and silicate to segregate on a planetary scale requires that at least the metal is molten (Stevenson, 1990). When the silicate (which has the higher melting temperature) is in a solid state, liquid metal can segregate by (a) grain-scale percolation, (b) the descent of km-size diapirs and/or (c) dyking (Fig. 1). On the other hand, when the silicate is also largely molten and present as a global-scale magma ocean, liquid metal can segregate extremely efficiently (Stevenson, 1990; Rubie et al., 2003). The heat that is required to produce melting originates from the decay of short-lived isotopes (especially $^{26}$Al) during the first 1-3 My of Solar System evolution and later from high-energy impacts between planetary bodies (Rubie et al., 2007; 2015a). The sinking of metal to the core also causes a temperature increase due to the conversion of potential energy to heat.

Many studies in recent years have concluded that core formation in the Earth involved extensive chemical equilibration between metal and silicate at high pressures (e.g. Li and Agree, 1996).

Here we discuss two mechanisms that are consistent with such equilibration, namely grain-scale percolation and segregation in a magma ocean. Because of slow diffusion rates in crystalline silicates (e.g. Holzapfel et al., 2005) and the large length scales involved, the diapir and dyking mechanisms result in insignificant chemical equilibration; these mechanisms are not discussed here but are reviewed by Rubie at al. (2007, 2015a). Note also that hybrid models have been proposed, such a combination of porous flow and diapirism (Ricard et al., 2009).

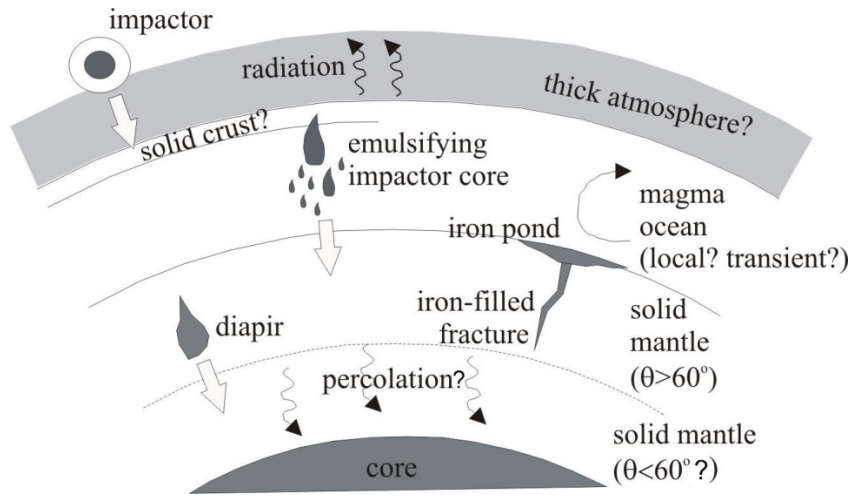

**Figure 1**. Summary of possible mechanisms by which liquid metal can segregate from silicate material during core formation. $\theta$ is the dihedral angle which controls grain scale percolation. Note that the feasibility of percolation in the lower mantle is uncertain (see text). (Courtesy of F. Nimmo.)

*2.1. Grain scale percolation*

Whether or not liquid metal can percolate through a polycrystalline silicate matrix depends on the dihedral angle $\theta$ between two solid-liquid boundaries where they intersect a solid-solid boundary at a triple junction (von Bargen and Waff, 1986; Stevenson, 1990). This dihedral angle is controlled by the solid-solid and solid-liquid interfacial energies. When the dihedral angle is less than 60°, the liquid metal is fully connected along grain edges and can percolate efficiently through the silicate matrix. When the dihedral angle exceeds 60°, the melt forms isolated pockets when the melt fraction is low and percolation is only possible when the volume fraction of melt exceeds some critical value that ranges from 2 to 6% for dihedral angles in the range 60-85° (see also Walte et al., 2007). By measuring dihedral angles in experimentally-sintered aggregates consisting of solid silicate + liquid Fe alloy, the feasibility of percolation as a core formation mechanism can be tested.

In general, experimental studies have been performed up to pressures of 25 GPa on samples in which a few volume percent of a liquid Fe alloy are contained in a polycrystalline aggregate of either olivine, ringwoodite or bridgmanite (silicate perovskite). In general, these studies have found that dihedral angles significantly exceed 60° and are little affected by pressure, temperature or the identity of the solid phase (e.g. Ballhaus and Ellis, 1996; Minarik et al., 1996; Shannon and Agee, 1996, 1998; Holzheid et al., 2000; Terasaki et al., 2005; 2007, 2008). A parameter that is of considerable importance is the oxygen and/or sulphur content of the liquid metal alloy. The dihedral angles decrease as the concentrations of these light elements increase (Terasaki et al., 2005). In contrast, Si and C dissolved in liquid Fe have little or no effect on dihedral angles (Mann et al., 2008, Li and Fei, 2014). Terasaki et al. (2008) showed that at pressures below 2-3 GPa, dihedral angles drop below 60° when the O + S content is high. Therefore, while percolation could be important during the differentiation of small bodies (planetesimals), it is unlikely to be important during core formation in larger planetary bodies.

Two studies have examined liquid Fe alloy interconnectivity in a $(Mg,Fe)SiO_3$-perovskite matrix using laser-heated diamond anvil cells (LH-DACs) up to 64 GPa. Using transmission electron microscopy to measure dihedral angles, Takafuji et al. (2004) found that dihedral angles apparently decrease from 94° at ~27 GPa/2400 K to 51° at ~47 GPa/3000 K, possibly as a consequence of increasing concentrations of Si and O in the metal. Shi et al. (2013) used in-situ X-ray tomography to measure dihedral angles and to generate 3-D images of metal distribution. They concluded that dihedral angles decrease from 72° at 25 GPa to 23° at 64 GPa and that the liquid Fe metal forms interconnected networks above 40-50 GPa. Although both studies concluded that percolation could have been an efficient segregation mechanism in the Earth's lower mantle, there are caveats to consider concerning the experimental technique. In particular, a sample that is pressurized in the LH-DAC is subjected to an extremely high differential stress that may have a major influence on microstructure, dihedral angles and connectivity. In fact, the tomographic images of Shi et al. (2013) suggest strongly that the interconnected metal in their samples is present along conjugate shear zones that result from high differential stress.

Furthermore, they reported dihedral angles as low as 12° at 52 GPa which raises the question as to whether textural equilibrium under hydrostatic stress conditions was attained.

Two additional factors to consider when discussing grain-scale percolation are the effects of (a) the presence of a small fraction of silicate melt and (b) deformation. Contrary perhaps to expectations, the presence of small fractions (e.g. 2-8 vol%) of silicate melt does not facilitate the percolation of Fe or FeS melt through largely-crystalline silicate. Thus high volume fractions of silicate melt are required for metallic liquids to segregate efficiently (Yoshino and Watson, 2005; Bagdassarov et al., 2009; Holzheid, 2013). Several studies have shown that deformation can enhance percolation in high dihedral angle systems especially when strain rates are high (e.g. Hustoft and Kohlstedt, 2006; Walte et al., 2011). However, the process is inefficient because small fractions of metallic liquid are always left stranded in the silicate matrix. In addition, it is unlikely that deformation at low strain rates (characteristic of mantle convection) enhances percolation (Walte et al., 2011).

Cerantola et al. (2015) have studied the effects of deformation on FeS segregation in a polycrystalline olivine matrix that also contains a small percentage of basaltic liquid. The results show that the presence of silicate liquid actually inhibits sulphide melt segregation by reducing its connectivity.

If grain scale segregation does occur (for example, during differentiation of planetesimals), it seems likely that chemical equilibration would be fast because of large surface to volume ratios, even though diffusion in solid silicates might be slow. To our knowledge, the extent and efficiency of such equilibration has never been modelled quantitatively.

*2.2. Metal-silicate segregation in magma oceans*

The late-stages of Earth accretion involved collisions with smaller planetesimals and embryos that culminated in the Moon-forming giant impact (Hartmann and Davis, 1975; Cameron and Ward, 1976; Chambers and Wetherill, 1998; Agnor et al., 1999). As well as delivering Fe metal to the Earth, such impacts provided sufficient energy to cause extensive melting and deep magma ocean formation (Tonks and Melosh, 1993; Rubie et al., 2007, 2015a). The delivery of energy is localized around the impact site and likely results in a roughly spherical melt pool (Fig. 2) that could extend to the core-mantle boundary in the case of a Mars-size impactor, as suggested for the Moon-forming event (Canup and Asphaug, 2001; Ćuk and Stewart, 2012).

Isostatic readjustment then results in the spreading out of the magma to form a global magma ocean hundreds of km deep (e.g. Fig. 1) but on a timescale that is uncertain (Reese and Solomatov, 2006). The lifetime of the global magma ocean is also very uncertain and could be short (e.g. several 1000 years) or long (~100 My) (especially in the case of a "shallow" magma ocean - which can have a basal pressure of up to 40 GPa) depending on the absence or presence of a dense insulating atmosphere (Abe, 1997; Solomatov, 2000). In the case of short-lived magma oceans, convection is turbulent with a Rayleigh number on the order of $10^{27}$-$10^{32}$ and convection velocities of several m/s (Solomatov, 2000; Rubie et al., 2003).

A magma ocean provides an environment in which metal-silicate segregation can occur rapidly and efficiently due to (a) the large density contrast between these materials and (b) the very low viscosity of ultramafic silicate liquids at high pressure (Liebske et al., 2005). Many impacting bodies are likely to have been already differentiated (Urey, 1955; Hevey and Sanders, 2006) and their metallic cores would plunge through the magma ocean. Within an end-member scenario, known as "core-merging", impactor cores remain intact and merge directly with the Earth's proto-core (e.g. Halliday, 2004). Within the other end-member scenario, impactor cores emulsify completely into small droplets as they sink through Earth's mantle (Stevenson, 1990; Rubie et al., 2003). In the first case there would be very limited chemical equilibration between metal and silicate at high pressure whereas in the second case, equilibration would be complete (Rubie et al., 2003). The mechanical behavior of impactor cores as they sink through a magma ocean thus determines the partitioning of siderophile elements between the core and mantle (as discussed below) and is also critical for estimating the timescale of core formation from W isotope anomalies (e.g. Nimmo et al., 2010).

Liquid metal sinking through silicate liquid tends to form droplets of a stable size that is controlled especially by the metal-silicate interfacial energy. Large metal blobs tend to break up as they sink due to mechanical instabilities whereas very small droplets coalesce to reduce interfacial energy (Rubie et al., 2003). The stable droplet size for typical magma ocean properties is ~1 cm diameter with a settling velocity of ~0.5 m/s (Rubie et al., 2003). The exact values depend on the Si-O-S content of the Fe alloy because this affects the metal-silicate interfacial energy (Terasaki et al., 2012).

Understanding the extent to which impactor cores emulsify into cm-size droplets in a magma ocean is a challenging problem. Emulsification is difficult to study numerically because relevant length scales vary over many orders of magnitude, from hundreds of kilometers to centimeters. Alternatively the problem can be approached through fluid dynamics experiments on analog liquids, although high impact velocities cannot then be taken into account. Currently it seems likely that planetesimal cores emulsify completely, whereas for much larger embryo cores the answer is still uncertain (Rubie et al., 2003; Olson and Weeraratne, 2008; Dahl and Stevenson, 2010; Deguen et al., 2011; Kendall and Melosh, 2014, 2015; Samuel, 2012; Landeau et al., 2014). Based on siderophile element partitioning and the Earth's mantle tungsten isotope anomaly, estimates of the fraction of core-forming metal that equilibrated with Earth's mantle include 36% (Rudge et al., 2010; note that this estimate increases to 53% when the new Hf/W measurements of König et al. (2011) are included), 30-80% (Nimmo et al., 2010) and 70-100% (Rubie et al., 2015b).

## 3. Geochemical models of core formation

### 3.1. Element partitioning and oxygen fugacity

The Earth's mantle is depleted in siderophile (metal-loving) elements, relative to chondritic and Solar System abundances, because of extraction into the core. The degree of depletion depends on the metal-silicate partition coefficient which, for element M, is described as:

$$D_{\mathrm{M}}^{met-sil} = \frac{C_{\mathrm{M}}^{met}}{C_{\mathrm{M}}^{sil}} \qquad (1)$$

where $C_{\mathrm{M}}^{met}$ and $C_{\mathrm{M}}^{sil}$ are the molar concentrations of M in metal and silicate respectively. Moderately siderophile elements (e.g. Ni, Co, W, Mo) are defined as having $D_{\mathrm{M}}^{met-sil}$ values $<10^4$ at 1 bar, highly siderophile elements (e.g. Ir, Pt, Pd, Ru) have $D_{\mathrm{M}}^{met-sil}$ values $>10^4$ and values for lithophile elements (e.g. Al, Ca, Mg) are <1. In general, partition coefficients depend on pressure, temperature and in some cases the compositions of the metal and silicate phases (Rubie et al., 2015a). In addition, oxygen fugacity ($f_{O2}$) is a critical controlling parameter as shown by the equilibrium:

$$M + (n/4)O_2 = MO_{n/2} \qquad (2)$$
metal         silicate liquid

where *n* is the valence of element M when dissolved in silicate liquid. At high oxygen fugacities, M is concentrated more in silicate liquid whereas at low $f_{O2}$ it is concentrated more in the metal. In addition, the dependence of partitioning on oxygen fugacity is strong in the case of high-valence elements and weak for low-valence elements. Oxygen fugacity is determined relative to the iron-wüstite buffer from the Fe content of metal and the FeO content of coexisting ferropericlase or silicate – both of which give comparable results (e.g. Asahara et al., 2004; Mann et al., 2009).

There have been numerous experimental studies of the dependence of siderophile element partitioning on pressure (*P*), temperature (*T*) and $f_{O2}$ with the aim of determining the conditions of metal-silicate equilibration during core formation (e.g. Rubie et al., 2015a, Table 3). In most studies the aim has been to reproduce core-mantle partition coefficients $D_M^{Core-Mantle} = C_M^{Core}/C_M^{Mantle}$ (e.g. Wood et al., 2006). Values of $C_M^{Mantle}$ are based on estimated primitive mantle concentrations (e.g., Palme and O'Neill, 2013) and $C_M^{Core}$ values are estimated by mass balance assuming a given Earth bulk composition (McDonough, 2003). Several types of core formation models have been fit to partitioning results, as reviewed below.

*3.2. Single stage core formation*

In the simplest and most commonly-applied model of core formation, chemical equilibration between the mantle and core at a single set of *P-T-* $f_{O2}$ conditions is assumed (e.g. Li and Agee, 1996; Corgne et al., 2009). This has led to a great variation of *P-T* estimates, ranging from 25 to 60 GPa and 2200 to 4200 K, mostly assuming an $f_{O2}$ about two log units below the iron-wüstite buffer (Rubie et al., 2015a, Table 3). A typical conclusion of such studies is that metal-silicate equilibration occurred at the base of a magma ocean at an equilibration pressure corresponding to the ocean's depth (i.e. in the range 700-1500 km). In this case, the equilibration temperature should lie close to the peridotite liquidus at the equilibration pressure. However, Wade and Wood (2005) found that $D_M^{core-mantle}$ values for Ni, Co, V, Mn and Si could be matched with an equilibration pressure of 40 GPa but only when the equilibration temperature exceeds the peridotite liquidus by ~700 K, which is physically implausible.

Righter (2011) determined a *P-T* estimate of 27-46 GPa and 3100-3600 K based on element partitioning data and argued that this represents the conditions of a final equilibration event that occurred at the culmination of Earth's accretion and growth. This seems to imply that a large fraction of the metal of the core equilibrated with a large fraction of the silicate of the mantle *in a single event* at pressure-temperatures conditions that correspond to a shallow mid-mantle depth.

*3.3. Continuous core formation and evolution of oxidation state*

In the "continuous" core formation model of Wade and Wood (2005), Earth accretion and the concurrent delivery of core-forming metal occurs in small steps of 1% mass (see also Wood et al., 2006; Wood et al., 2008). Each batch of metal equilibrates with the silicate magma ocean at its base, the depth of which increases as the Earth grows. Thus metal-silicate equilibration pressures and temperatures (the latter defined by the peridotite liquidus) increase during accretion. In order to reproduce core-mantle partition coefficients of Ni, Co, Cr, V, Nb and Si, a magma ocean thickness corresponding to 35% of mantle depth is required and *P-T* equilibration conditions reach a maximum of 40-50 GPa and ~3250 K at the end of accretion. However, reproduction of mantle concentrations was not possible when $f_{O2}$ remains constant and, instead, conditions need to become increasing oxidizing, by ~2 log units, during accretion (see also Rubie et al., 2011).

For $f_{O2}$ to increase significantly during accretion requires the FeO content of the mantle to increase (e.g. from <1 to 8 wt%), for which there are two viable mechanisms (Rubie et al., 2011). First, when Si partitions into the core, the FeO content of the mantle increases by the reaction:

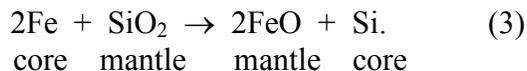

$$2Fe\ +\ SiO_2\ \rightarrow\ 2FeO\ +\ Si. \qquad (3)$$
$$\text{core}\quad\text{mantle}\quad\ \text{mantle}\quad\text{core}$$

Thus for every mole of Si that partitions into the core, two moles of FeO are added to the mantle, which means that this mechanism is very effective at oxidizing the mantle. Second, the accretion of relatively oxidized material during the later stages of accretion can also significantly increase the mantle FeO content. Note that the "oxygen pump" or "self-oxidation" mechanism proposed by Wade and Wood (2005) increases $Fe^{3+}$ but not the FeO content of the mantle: it is therefore not a viable oxidation mechanism in the present context (Rubie et al., 2015a).

It has also been proposed that core formation may occur under initially oxidizing conditions (Rubie et al., 2004; Siebert et al., 2013). This model is based on the Earth's mantle/magma ocean initially containing ~20 wt% FeO. The resulting high $f_{O2}$ causes FeO to partition into the core so that the mantle FeO is progressively reduced during accretion to its current value of 8 wt%. However, if a small amount (e.g. 2-3 wt%) of Si also partitions into the core, which is inevitable at high temperatures (Siebert et al., 2013), this model fails based on mass balance. This is because the initial ~20 wt% FeO is reduced only slightly during accretion (to 17-18 wt%) because of the production of FeO by reaction (3) (Rubie et al., 2015b).

*3.4. Multistage core formation*

The Earth accreted through a series of high-energy impacts with smaller bodies consisting of km- to multi-km-size planetesimals and Moon- to Mars-size embryos (e.g. Chambers and Wetherill, 1998). Such impacts, as well as delivering energy that caused extensive melting, added Fe-rich metal which segregated to the Earth's proto-core. Thus core formation was multistage and was an integral part of the accretion process.

A preliminary model of multistage core formation is based on an idealized accretion scenario in which the Earth accretes through impacts with differentiated bodies that have a mass ~10% of Earth's mass at the time of each collision (Rubie et al., 2011). The metal of the impacting bodies equilibrates, partially or completely, in a magma ocean at a pressure that is a constant fraction of the Earth's core-mantle boundary pressure and at a temperature close to the corresponding peridotite liquidus. Thus, as in the model of continuous core formation, metal-silicate equilibration pressures increase as the Earth grows in size. In contrast to previous studies, this model is not based on assumptions about oxygen fugacity and its evolution. Instead, the bulk compositions of the accreting bodies are defined in terms of non-volatile elements, which are assumed to be present mostly in Solar System (CI chondritic) relative abundances. The oxygen content is the main compositional variable that enables a wide range of compositions between two extreme end members to be defined: at low oxygen content, all Fe is present as metal whereas at high oxygen content all Fe is present as FeO in the silicate.

The compositions of equilibrated metal and silicate liquids, at a given *P-T*, are expressed as:

$$[(FeO)_x\,(NiO)_y\,(SiO_2)_z\,(Mg_u\,Al_m\,Ca_n)O] + [Fe_a\,Ni_b\,Si_c\,O_d]. \qquad (4)$$
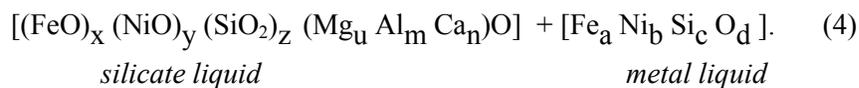
*silicate liquid*                                      *metal liquid*

The indices u, m and n are determined from the bulk composition alone because Mg, Al and Ca do not partition into the metal. The other 7 indices, x, y, z, a, b, c and d, are determined by simultaneously solving 4 mass balance equations (for Fe, Si, O and Ni), 2 partitioning expressions for Si and Ni, and a model of oxygen partitioning (Frost et al., 2010) by an iterative process that is described in detail by Rubie et al. (2011, supplementary data). Trace elements have little effect on the mass balance and concentrations in metal and silicate are based on partitioning alone. Using this approach, the core of each impactor is equilibrated (fully or partially) with the magma ocean at high *P-T* and the resulting metal is added to the proto-core. Thus the evolution of mantle and core compositions is modeled throughout Earth's accretion (Rubie et al., 2011, Fig. 4).

For simplicity, two bulk compositions are used in this model of Earth accretion: (1) A highly reduced composition in which 99.9% of Fe and ~20% of available Si are present as metal, and (2) A relatively oxidized composition in which ~60% of Fe is present as metal and ~40% as FeO (the initial metal contents of these two compositions are 36 wt% and 20 wt%, respectively). These compositional parameters, together with equilibration pressures and extent of metal re-equilibration, are refined by a least squares minimization in order to fit the final mantle concentrations of FeO, $SiO_2$, Ni, Co, W, Nb, V, Ta and Cr to those of the Earth's primitive mantle. Best results are obtained when the initial 60-70% of the Earth accretes from the reduced composition and the final 30-40% from the more oxidized material, with equilibration pressures 60-70% of core-mantle boundary pressures at the time of each impact. In addition, during the final few impacts, only a limited fraction of the metal equilibrates with silicate. Note that, apart from the need for high-pressure metal-silicate equilibration, this model has similarities to early models of heterogeneous accretion (Wänke, 1981; O'Neill, 1991).

*3.5. Combined accretion and core-mantle differentiation model*

The multistage core formation model described above has recently been extended by combining it with N-body accretion simulations (Rubie et al., 2015b). The latter study concentrates on "Grand Tack" accretion models because of their success in reproducing the masses and orbital characteristics of the terrestrial planets and especially Earth and Mars (Walsh et al., 2011; O'Brien et al., 2014; Jacobson and Morbidelli, 2014). Using the mass balance/partitioning approach described above, the compositions of primitive bodies in the solar nebula are defined in terms of oxidation state and water content as a function of heliocentric distance by least squares regressions. This is done by adjusting the fitting parameters in order to produce an Earth-like planet with a mantle composition identical to (or close to) that of the Earth's primitive mantle. The only composition-distance model that provides acceptable results involves bodies close to the Sun (<0.9-1.2 AU) having a highly reduced composition and those from further out being increasingly oxidized (Fig. 3a). Beyond the giant planets, all bodies are completely oxidized (i.e.

with no metallic cores) and contain 20 wt% $H_2O$, which results in ~1000 ppm $H_2O$ in the Earth's mantle. Note that in the Grand Tack model, the C-complex asteroids, which are thought to be carbonaceous chondrite parent bodies containing water that matches that of the Earth (Morbidelli et al., 2000; Alexander et al., 2012), are delivered to the Earth and the outer Main Belt from their initial locations at the inner edge of an extended outer disk (i.e. beyond Saturn) by the outward migration of the giant planets (Walsh et al., 2011; O'Brien et al., 2014). The bodies originally located between the snow line and 6-8 AU were swept up by Jupiter and Saturn and were not accreted to the terrestrial planets. (The snow line is the minimum distance from the Sun in the protoplanetary disk at which temperatures were low enough for water ice to condense. It was likely located at ~3 AU.)

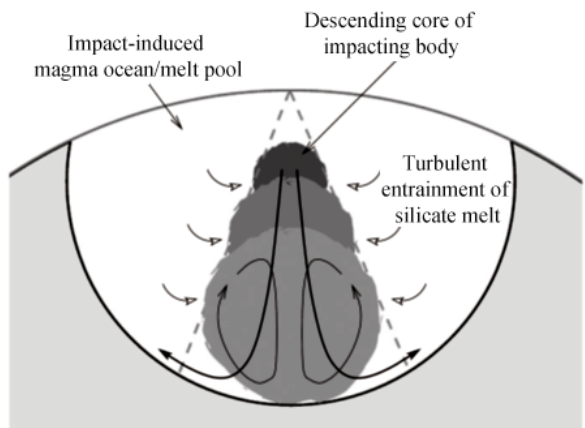

**Figure 2**. Fluid dynamical model of the descent of an impactor's iron core through an impact-induced semi-spherical magma ocean/melt pool. As the core sinks, silicate liquid is turbulently entrained in a descending metal-silicate plume which broadens with depth. Only the silicate liquid that is entrained in the plume equilibrates chemically with the metal. After Deguen et al. (2011).

Hundreds of impacts and associated core-formation events are simulated for all embryos, and thus the final terrestrial planets, simultaneously. For the first time, metallic cores of impacting bodies are modeled to only equilibrate with a small fraction of the target's mantle/magma ocean (Fig. 2); this is in contrast to all previous studies in which it has been assumed that the entire magma ocean equilibrates with the metal. One consequence is that metal-silicate equilibration pressures need to be relatively high (Rubie et al., 2015b).

Earth's mantle composition is perfectly reproduced in both simulations shown in Fig. 3b,c. The Martian mantle composition is correctly predicted in the simulation shown in Fig. 3b but is much too FeO-poor in the simulation of Fig. 3c – this is because the embryo that formed Mars originated too close to the Sun, and therefore under conditions that are too reducing in simulation 2:1-0.25-10A (for further discussion see Rubie et al., 2015b). Mercury's FeO-poor mantle is not reproduced in either simulation because, although it starts with an appropriately-reduced composition, Mercury subsequently accretes too much oxidized material in both simulations.

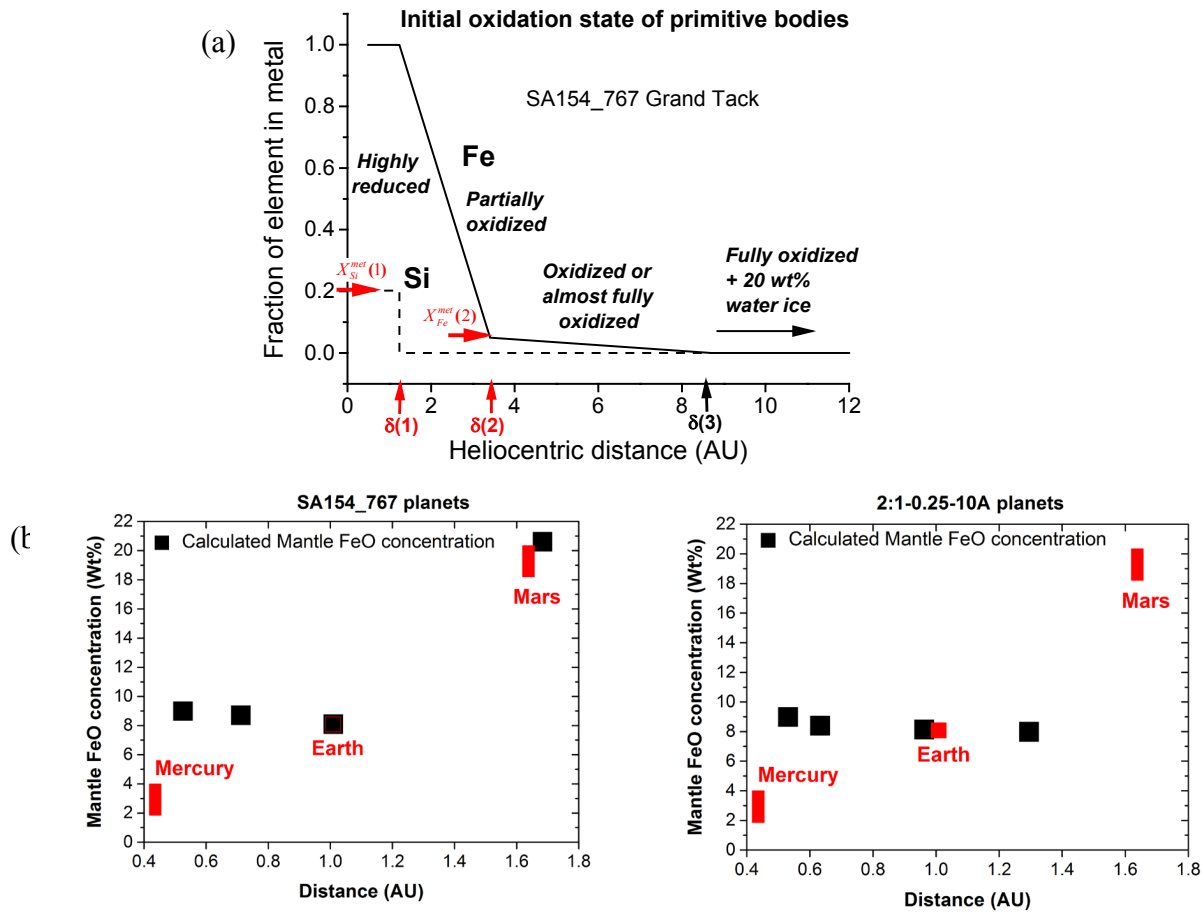

**Figure 3**. Results of combined Grand Tack N-body accretion/core formation models. (a) Best-fit composition-distance model for primitive bodies at heliocentric distances from 0.7 to 12 AU for N-body simulation SA154_767. Oxidation state is defined by the fractions of total Fe and Si that are initially present as metal. For example, in the highly-reduced composition, 99.9% of bulk Fe and 20% of bulk Si are present as metal  The four parameters labelled in red, together with equilibration pressure, were refined by least squares minimization to fit Earth's mantle composition. (b) and (c) Calculated FeO contents of the mantles of Mercury, Venus, Earth and Mars analogs in simulations SA154_767 and 2:1-0.25-10A respectively. Actual mantle compositions of Mercury, Earth and Mars are shown as red symbols.

The ability to predict mantle and core compositions of Mercury, Venus and Mars, in addition to that of the Earth, adds a new constraint to both accretion and core formation models. For example, in terms of reproducing the composition of the Martian mantle, the accretion model of Fig. 3b is more successful than that of Fig. 3c.

## 4. Conclusions and outlook

Accretion and core-mantle differentiation of the terrestrial planets can now be modeled by combining astrophysical simulations of planetary accretion with geochemical models of core formation (Rubie et al., 2015b). Such models are becoming increasing sophisticated and realistic in terms of incorporating new constraints on both physical and chemical processes. The approach enables the compositions and mass ratios of the cores and mantles of all the terrestrial planets to be modeled simultaneously. By fitting the mantle compositions of model Earth-like planets (e.g. located at ~1 AU and of ~1 Earth mass), fitting parameters can be defined by least squares minimization. In addition, the bulk compositions of primitive bodies at heliocentric distances ranging from from ~0.7 to 10 AU can be defined in terms of oxidation state and water content. These combined models can place new constraints on our understanding of both planetary accretion and core-mantle differentiation processes.

Currently only non-volatile elements and $H_2O$ have been considered in the combined model. However, there is scope for also including a range of other elements, such as S and other volatile elements, highly siderophile elements, and both stable and radiogenic isotopes.

Currently there are a number of caveats that must be mentioned (see Rubie et al., 2015b, for a detailed discussion). For example, element partitioning data are often extrapolated from <25 GPa to pressures as high as 80 GPa. In order to reduce the uncertainties, careful but challenging experimental studies using the diamond anvil cell are required. For example, the partitioning of Si, O, Ni, Co, V and Cr between metal and silicate has been studied recently to 100 GPa and 5500 K (Fischer et al., submitted). In addition, partition coefficients are often a function of the concentrations of elements such as Si, O and S in the liquid metal, although for most of the elements considered so far, the effects are small (e.g. Tuff et al., 2011). Finally, accretion models are currently based on the assumption of 100% efficiency, whereas in reality, material will often be lost during accretional impacts.

**Acknowledgements**

D.C.R. and S.A.J. were supported by the European Research Council (ERC) Advanced Grant "ACCRETE" (contract number 290568). We thank Francis Nimmo for kindly providing Figure 1, David O'Brien, Alessandro Morbidelli, Herbert Palme and Ed Young for discussions, and Kevin Righter and an anonymous referee for helpful reviews.